\documentclass[aps,prb,twocolumn,notitlepage,amsfonts,citeautoscript,superscriptaddress]{revtex4-2}
\usepackage{xr}
\usepackage[caption=false]{subfig}
\usepackage{amsmath,amssymb,mathrsfs}
\usepackage{latexsym}
\usepackage{natbib}
\usepackage{graphicx}
\usepackage{amsfonts}
\usepackage[dvipsnames]{xcolor}
\usepackage{bm}
\usepackage{url}
\usepackage{soul}
\usepackage{cancel}
\usepackage{hyperref}
\usepackage{cleveref}
\usepackage{lipsum}
\usepackage[version=4]{mhchem}

\usepackage{color}

\usepackage{changes}
\setcitestyle{super}

\hypersetup{
    colorlinks = true,
    linkcolor = MidnightBlue,
    citecolor = MidnightBlue,
    linkbordercolor = {white},
    urlcolor = RoyalBlue
}

\crefname{equation}{Eq.}{Eqs.}
\Crefname{equation}{Equation}{Equations}
\crefname{figure}{Fig.}{Figs.}
\Crefname{figure}{Figure}{Figures}
\crefname{section}{Sec.}{Secs.}
\Crefname{section}{Section}{Sections}
\crefname{appendix}{Appendix}{Apps.}
\Crefname{appendix}{Appendix}{Apps.}
\crefname{paragraph}{Sec.}{Secs.}
\crefname{table}{Table}{Tables}

\newcommand{\textalert}[1]{}

\newcommand{\ket}[1]{\left|#1\right\rangle}
\newcommand{\bra}[1]{\left\langle#1\right|}

\newcommand{\braket}[2]{\bigl\langle#1\bigl|\bigr.#2\bigr\rangle}
\newcommand{\ketbra}[2]{\bigl|\bigr.#1\bigr\rangle\bigl\langle#2\bigr|}

\def\ie{\emph{i.e.},\ }
\def\eg{\emph{e.g.}\ }

\definechangesauthor[name=Ivano, color=Magenta]{ita}
\definechangesauthor[name=Alberto, color=Orange]{ab}
\definechangesauthor[name=Anthony, color=RoyalBlue]{ag}

\begin{document}

\title{Non-adiabatic quantum dynamics with fermionic subspace-expansion algorithms on quantum computers}

\author{Anthony Gandon}
\affiliation{IBM Quantum, IBM Research - Zurich, Säumerstrasse 4, 8803 Rüschlikon, Switzerland}
\affiliation{Institute for Theoretical Physics, ETH Zürich, Wolfgang-Pauli-Str. 27, 8093 Zürich, Switzerland}

\author{Alberto Baiardi}
\affiliation{IBM Quantum, IBM Research - Zurich, Säumerstrasse 4, 8803 Rüschlikon, Switzerland}

\author{Pauline Ollitrault}
\affiliation{QC Ware, Palo Alto, USA, and Paris, France}

\author{Ivano Tavernelli}
\affiliation{IBM Quantum, IBM Research - Zurich, Säumerstrasse 4, 8803 Rüschlikon, Switzerland}

\begin{abstract}
We introduce a novel computational framework for excited-states molecular quantum dynamics simulations driven by quantum computing-based electronic-structure calculations.
This framework leverages the fewest-switches surface-hopping method for simulating the nuclear dynamics, and calculates the required excited-state transition properties with different flavors of the quantum subspace expansion and quantum equation-of-motion algorithms.
We apply our method to simulate the collision reaction between a hydrogen atom and a hydrogen molecule.
For this system, we critically compare the accuracy and efficiency of different quantum subspace expansion and equation-of-motion algorithms and show that only methods that can capture both weak and strong electron correlation effects can properly describe the non-adiabatic effects that tune the reactive event.
\end{abstract}

\maketitle


\section{Introduction}
\label[section]{introduction}

Quantum computing is emerging as a new computational paradigm with the potential of transforming algorithms for electronic structure calculations performed on classical hardware\cite{Cao2019_QuantumChemistryAge, Bauer2020_QuantumAlgorithms}. 
With currently available noisy quantum hardware, it is nonetheless quite improbable that large and complex physics and chemical systems will be entirely simulated on quantum devices. 
Most likely, mixed-quantum classical algorithms, where only a part of the full problem is solved on the quantum device, will play an important role for near-term quantum simulations. 
This class of approaches includes embedding schemes, in which the electronic system is partitioned into an `easier' component that can be successfully described with approximate, classically-efficient methods (\eg Density Functional Theory), and a `harder' subsystem (\eg a strongly correlated fragment) that is solved more accurately on a quantum computer\cite{Rossmannek2021_QuantumHFDFTEmbedding,Ralli2022_ScalableApproachQuantum,Rossmannek2023_QuantumEmbeddingMethod}. Additionally, error mitigation techniques\cite{Temme2017_ErrorMitigation} remain indispensable for achieving the required level of accuracy. For a successful implementation of such schemes, an efficient integration of quantum and classical algorithms is crucial.

The Variational Quantum Eigensolver (VQE) has appeared in the last decade as a promising hybrid classical-quantum algorithm to run quantum chemical calculations on quantum computers before the thresholds for fault tolerance are reached~\cite{Peruzzo2014_VQE,Kandala2017_HardwareEfficient-VQE,Tilly2022_VQE-Review}.
The VQE optimizes classically the energy of a given parametrized ansatz state, $\ket{\psi(\theta)} = U(\theta)\ket{\psi_\mathrm{init}}$, with respect to a fixed Hamiltonian $\hat{H}$ by evaluating the energy cost function on a quantum device.
The variational principle guarantees that the energy landscape will be bounded from below by the exact ground state energy of $\hat{H}$ independently on the ansatz $U(\theta)$. In particular, increasingly more refined ans\"atze, ranging from hardware-inspired\cite{Kandala2017_HardwareEfficient-VQE} to physically motivated\cite{Romero2018_UCC-VQE,leeGeneralizedUnitaryCoupled2019,Sokolov2020_UCC-StrongCorrelation,grimsleyAdaptiveVariationalAlgorithm2019b,tangQubitADAPTVQEAdaptiveAlgorithm2021a,anselmettiLocalExpressiveQuantumnumberpreserving2021b,burtonExactElectronicStates2023,burtonAccurateGateefficientQuantum2024} ones, can be realized with relatively shallow quantum circuits.

Most of VQE applications focus on ground-state energy calculation, although recent works have investigated the estimation of excitation energies for general Hamiltonians with algorithms tailored to near-term devices. Several strategies have been proposed, inspired by techniques developed for classical quantum-chemical simulations:
\begin{enumerate}
    \item Symmetry-adapted approaches,\cite{Zhang2021_VQE-Symmetry,Cao2022_PointGroup-Symmetry,Goings2023_VQE-Benzene-Symmetry} which rely on \emph{ansatz} structures preserving certain quantum numbers. Ground states in well-chosen symmetry sectors correspond to excited states of the full Hilbert space. This approach requires minor modifications to the VQE algorithm, but it is limited to highly-symmetric molecule with as many symmetries as target low-energy excited states.
    \item Penalty-based optimization,\cite{Izmaylov2019_Penalty-VQE,Higgott2019_Deflation,Xie2022_Orthogonal-ES-VQE,Carobene2023_Penalty-VQE} in which a term is added to the Hamiltonian to either penalize states with quantum numbers different than the target ones or to impose orthogonality other previously calculated ground or excited states.
    \item Subspace expansion approaches,\cite{Urbanek2020_QuantumSubspace,Stair2020_Krylov,McClean2020_qSE,ollitrault_quantum_2020,Cortes2022_QuantumKrylov, Reinholdt2024SubspaceMethodsSimulation} in which the excited states are expanded as a linear combination of intermediary states, often built on top of the ground state approximate wave function.
\end{enumerate}

The extension of variational approaches to excited-state calculations represents a crucial step to broaden their application range. In fact, while the electronic ground state of many classes of molecular systems are (heuristically) known to display weak correlation (e.g., for organic molecules), strong correlation effects are omnipresent in excited states, even for simple molecular systems. If well-established classically efficient quantum-chemical methods exist for weakly-correlated molecules, simulating molecules with many strongly-correlated electrons remains a major challenge for classical computing. 

In the present work we focus on subspace expansion approaches. More specifically, we consider chemically motivated fermionic subspaces for describing the low-energy spectrum of chemical systems, where the intermediary states are built applying fermionic excitations on the variational VQE ground state approximation\cite{McClean2020_qSE}. We study and compare different flavors of fermionic subspace expansion, not only in terms of complexity and accuracy of the energy estimates, but also in the practical context of non-adiabatic molecular dynamics (MD). To this end, we present a complete framework for \emph{ab-initio} ground and excited state MD calculations, where all relevant observables for the dynamics - excitation energies, nuclear energy gradients, non-adiabatic couplings - are estimated on the quantum processor. We apply this framework to the simple, yet non-trivial system of a dihydrogen molecule-hydrogen atom collision process.

This paper is organized as follows: we introduce in \cref{simulationTarget} the prototypical chemical system for molecular dynamics that we will investigate in this work. In \cref{fermionicSubspaceExpansion} we summarize the main aspects of the fermionic subspace expansion theory and its application to the calculation of electronic excited-state properties. We discuss in \cref{qeomDefinitionApplicability} theoretical aspects of different formulations of fermionic expansion for estimating these properties, and illustrate our different findings on the target system. Motivated by our goal of simulating molecular dynamics, we compare in \cref{applicationToMolecularDynamics} these methods for the calculation of population transfer during a scattering event, relying on all previously estimated static observables.


\section{Simulation target}
\label[section]{simulationTarget}

The system we study in this work consists of a dihydrogen molecule-hydrogen atom complex.
We align the dihydrogen molecule along the y-axis and place its center of mass at the origin.
We then simulate the scattering event resulting from a single hydrogen atom approaching with an almost perpendicular incident angle.
We chose this system because, despite its simplicity, it exhibits an avoided crossing between its ground and first excited state close to the equilateral geometry. In the vicinity of the avoided crossing, we expect non-adiabatic contributions to strongly influence the dynamics and to increase the ground state correlation. These contributions are notably hard to capture with conventional quantum-classical methods. Following the simulations reported in Ref.~\citenum{Tavernelli2009_TDDFT-NonAdiabatic} based on time-dependent density functional theory, we avoid the singularity at the conical intersection by moving slightly away from the exact equilateral geometry and by fixing the incident angle to $88^{\circ}$.

The molecular complex consists of 3 nuclei and 3 electrons and its electronic structure Hamiltonian expressed in the minimal STO-3G basis set (comprising 6 spin-orbitals) is
\begin{equation}
  \hat{H} = \sum_{pq}^6 h_{pq} \hat{a}_p^\dagger \hat{a}_q 
    + \frac{1}{2} \sum_{pqrs}^6 \langle p q | r s \rangle \hat{a}_p^\dagger \hat{a}_q^\dagger \hat{a}_s \hat{a}_r \, ,
  \label{eq:SQHamiltonian}
\end{equation}
where $h_{pq}$ and $\langle pq | rs \rangle$ are the conventional one- and two-electron integrals, and $\hat{a}^\dag$ and $\hat{a}$ are fermionic orbital creation and annihilation operators.\cite{Barkoutsos2018_ParticleHole}
These operators can be mapped to Pauli operators acting on 6 qubits through the Jordan-Wigner transformation\cite{Jordan1928_JW-Transforamtion}. We further reduce the number of qubits to 4 using the qubit tapering introduced in Ref.~\citenum{Bravyi2017_Tapering} based on symmetries of the Hamiltonian.
The resulting Hamiltonian, mapped in the spin-$1/2$ basis that is suitable for quantum computers, is a weighted sum of Pauli terms $\hat{P}_i \in \{I,X,Y,Z\}^{\otimes 4}$
\begin{equation}
    \hat{H}_p = \sum_i c_i \hat{P}_i \, .
    \label{eq:SumOfPauli}
\end{equation}

We obtain our reference values by exact diagonalization of this tapered Hamiltonian in the Hilbert space of dimension $2^4$. Note that his classical diagonalization is only possible because of the small system size.

We study the molecular dynamics of \ce{H2}-\ce{H} within the Born-Oppenheimer approximation.
We first sample the nuclear geometry based on Wigner sampling\cite{Wigner1932_DistributionFunction} and, for each nuclear configuration, we optimize the electronic ground state on a quantum device emulator.
We use an ADAPT-VQE\cite{Grimsley2019_AdaptVQE} optimization based on the Unitary Coupled-Cluster Singles and Doubles (UCCSD) ansatz to approximate the ground state $\ket{\psi_{vqe}}$ starting from the mean-field reference Hartree-Fock state.
All the excited-state computations are based on measurements of observables on the optimized ground state, which do not increase the circuit depth.
The solution of the eigenvalue problems associated with the resulting subspace Hamiltonians (see below for additional details about how the subspace Hamiltonians are constructed) can be computed on a classical computer as the typical size of the expansion subspace $d$ is small compared to the original Hilbert space.

The library \texttt{Qiskit}\cite{Qiskit} was used for the optimization of our ansatz circuits as well as the evaluation of quantum observables. We then interface the emulated quantum-computing calculations with the classical Tully's fewest switches surface hopping\cite{Tully1990_FSSH} Python code \texttt{Mudslide}\cite{Mudslide} to simulate \emph{ab-initio} non-adiabatic  molecular dynamics on the sampled Potential Energy Surfaces (PESs) of the system.

\section{Fermionic Subspace Expansion for excited state properties}
\label[section]{fermionicSubspaceExpansion}

In this section, we discuss the basic theory of quantum subspace expansion, focusing on aspects relevant for the calculation of excited-state properties and for their applications in MD simulations.

\subsection{Fermionic subspace expansion}

Subspace expansion methods approximate certain excited states of a given Hamiltonian $\hat{H}$ by diagonalizing a projection of this operator in a small, well-chosen, subspace. In this work, we define the expansion subspace from a set of intermediary states
\begin{equation}
  \{\ket{\psi_\alpha}~|~\alpha=1,\cdots d\} \, .
  \label{eq:SubspaceDefinition}
\end{equation}
In order for subspace expansion methods to be efficient, these states should have a large overlap with the true excited states, and the size $d$ should scale sub-exponentially with the full Hilbert space.
Diagonalizing the projected Hamiltonian within this subspace yields $d$ approximated excited states as linear combinations of the intermediary states $\ket{\psi_\alpha}$. 
Note that although we mainly use the term ``approximation'' when the quantities become exact in the limit of a full Hilbert space description, and the term ``estimation'' in the statistical sense, the meaning of the two terms should be easy to understand based on the context.
In practice, projecting the Hamiltonian in the subspace spanned by these vectors requires calculating all matrix elements $\bra{\psi_\alpha} \hat{H} \ket{\psi_\beta}$ and $\braket{\psi_\alpha}{\psi_\beta}$.
In the context of quantum algorithms, this operation is assumed to be classically hard and is carried out on the quantum device.

In this work we consider expansion sets \cite{McClean2020_qSE, ollitrault_quantum_2020} constructed from a ground state approximation $\ket{\psi_{vqe}}$, here the result of a previous VQE-like optimization with an ansatz circuit $U(\theta)$ applied on the Hartree-Fock reference state $\ket{\mathrm{HF}}$, and from operators $\hat{E}_\alpha$ admitting a compact second-quantized representation
\begin{equation}
  \{ \ket{\psi_{vqe}},~\hat{E}_\alpha \ket{\psi_{vqe}}~|~\alpha=1,2,\cdots d\} \, .
  \label{eq:SubspaceVQE}
\end{equation}

Then, the calculation of the matrix elements introduced above can be reformulated as estimating $\langle \hat{A} \rangle_{vqe} = \langle \psi_{vqe} \vert \hat{A} \vert \psi_{vqe} \rangle$ the expectation value of various observables $\hat{A}$ on the approximated ground state.


\subsection{Diagonalization of the subspace Hamiltonian}

The quantum subspace expansion (qSE) proposed in Ref.~\citenum{mcclean_hybrid_2017} expands the excited-state wave function into a subspace generated by applying fermionic excitation operators of a fixed, given rank onto the ground-state wave function $\ket{\psi_{vqe}}$.
The fermionic excitation operators can be expressed as $a_m^\dag a_n^\dag...a_i a_j...$ and they represent the excitation of electrons from the occupied orbitals $i, j,\dots$ to the virtual orbitals $m, n,\dots$.
Including single $a^\dag_m a_i$ and double $a^\dag_m a^\dag_n a_i a_j$ electron excitations was shown to yield accurate approximations of the lowest lying excited energies of small molecules such as $\ce{H2}$, $\ce{LiH}$, and $\ce{H2O}$. \cite{Colless2018_qSE-ES,McClean2020_qSE}

Expressing these excitations as operators $\hat{E}_\alpha$ where $\alpha\in\{(m,i),...\}\cup\{(m,n,i,j), ...\}$, we construct the projected Hamiltonian $\bm{H}_{qse}$ and the overlap matrix $\bm{S}_{qse}$ in the subspace $\mathcal{E}_{qse} = \{\ket{\psi_{vqe}}, \hat{E}_\alpha\ket{\psi_{vqe}}~\forall \alpha\}$ as
\begin{align}
    \bm{H}_{qse} & =
    \begin{bmatrix}
        \langle \hat{H} \rangle_{vqe} & 
        \langle \hat{H} \hat{E}_\beta \rangle_{vqe} \\
        \langle \hat{E}_\alpha^\dag \hat{H} \rangle_{vqe} & 
        \langle \hat{E}_\alpha^\dag \hat{H} \hat{E}_\beta \rangle_{vqe}
    \end{bmatrix}, \\
    \bm{S}_{qse} & =
    \begin{bmatrix}
        \langle \hat{I}\rangle_{vqe} & 
        \langle \hat{E}_\beta \rangle_{vqe} \\
        \langle \hat{E}_\alpha^\dag \rangle_{vqe} & 
        \langle \hat{E}_\alpha^\dag \hat{E}_\beta \rangle_{vqe}
    \end{bmatrix},
\end{align}
where each matrix element corresponds to the evaluation of a different observable on the quantum circuit preparing the approximate ground state.
Note that the observables are mapped to the qubit space using the Jordan-Wigner transformation and the qubit tapering technique described above.
The approximated excited states defined in this subspace are then the solutions of a projected eigenvalue problem
\begin{equation}
  \label[equation]{qse_pseudoeigenvalue_problem}
  \overbrace{\hat{H}\ket{k} = \epsilon_k \ket{k}}^{exact} \longrightarrow \overbrace{\bm{H}_{qse} X_k = \bm{S}_{qse} X_k \tilde{\epsilon}_k}^{\text{in}~\mathcal{E}_{qse}} \, .
\end{equation}

Note that we write the \emph{exact} eigenstates as  $\ket{k}$ to distinguish them from their approximate counterparts $\hat{O}_k\ket{\psi_{vqe}} = X_k^0 \hat{I} \ket{\psi_{vqe}} + \sum_\alpha X_k^\alpha \hat{E}_\alpha \ket{\psi_{vqe}}$.
The eigenenergies $\tilde{\epsilon}_k$ are real because $\bm{H}_{qse}$ and $\bm{S}_{qse}$ are hermitian.
From the perspective of optimization, solving \cref{qse_pseudoeigenvalue_problem} for the $k$-th excited state can be understood as minimizing the energy landscape
\begin{equation}
    \label[equation]{qse_cost_function}
    f_{qse}(X_k) = \frac{\bra{\psi_{vqe}}\hat{O}_k^\dag \hat{H} \hat{O}_k \ket{\psi_{vqe}}}{\bra{\psi_{vqe}}\hat{O}_k^\dag \hat{O}_k \ket{\psi_{vqe}}} = \frac{X_k^\dag \bm{H}_{qse} X_k}{X_k^\dag \bm{S}_{qse} X_k}
\end{equation}
with respect to the coefficients $X_k$ or, equivalently, to the trial operator $O_k$, with orthogonality constraints
\begin{equation}
    \forall \, i<k,~\bra{\psi_{vqe}}\hat{O}_k^\dag \hat{O}_i\ket{\psi_{vqe}} = X_k^\dag S_{qse} X_i = 0 \, .
    \label[equation]{qse_orthogonality_constraints}
\end{equation}

The point of view of optimization will be used in later sections. However, to avoid confusion, we note that the subspace expansion approaches in truncated subspaces are never strictly variational for the excited states. Indeed, while \cref{qse_cost_function} with the constraints of \cref{qse_orthogonality_constraints} can be associated with a variational principle, the corresponding global minima will not be the $k$-th exact excited state because the constraints are defined from the approximate states $\hat{O}_i\ket{\psi_{vqe}}$ instead of the exact excited states $\ket{i}$. 

\subsection{Including fermionic excitations and de-excitations in the subspace}

For ground state wave functions of systems with non-negligible correlation, including only excitation operators in the operator pool -- an approximation usually referred to as Tamm-Dancoff approximation (TDA) -- may not be sufficient for accurately representing the excited-state wave functions. In these cases, de-excitation operators may contribute significantly to the excited-state wave functions. De-excitations can be added to the generators of the expansion subspace by building $\mathcal{F}_{qse} = \{\ket{\psi_{vqe}}\} \cup \{ \hat{E}_\alpha \ket{\psi_{vqe}}~\forall \alpha \} \cup \{ \hat{E}_\alpha^\dag \ket{\psi_{vqe}}~\forall \alpha \}$. This yields a larger and, therefore, more precise representation of the Hamiltonian at the price of extra measurements needed to construct it,
\begin{align}
    \bm{H}_{qse} & =
    \begin{bmatrix}
        \langle \hat{H} \rangle_{vqe} & \langle \hat{H} \hat{E}_\beta \rangle_{vqe} & \langle \hat{H} \hat{E}^\dag_\beta \rangle_{vqe} \\
        \langle \hat{E}_\alpha^\dag \hat{H} \rangle_{vqe} & \langle \hat{E}_\alpha^\dag \hat{H} \hat{E}_\beta \rangle_{vqe} & \langle \hat{E}_\alpha^\dag \hat{H} \hat{E}^\dag_\beta \rangle_{vqe}\\
        \langle \hat{E}_\alpha \hat{H} \rangle_{vqe} & \langle \hat{E}_\alpha \hat{H} \hat{E}_\beta \rangle_{vqe} & \langle \hat{E}_\alpha \hat{H} \hat{E}^\dag_\beta \rangle_{vqe}
    \end{bmatrix}, \label[equation]{extendedHamiltonian}\\
    \bm{S}_{qse} & =
    \begin{bmatrix}
        \langle \hat{I} \rangle_{vqe} & \langle \hat{E}_\beta \rangle_{vqe} & \langle \hat{E}^\dag_\beta \rangle_{vqe} \\
        \langle \hat{E}_\alpha^\dag \rangle_{vqe} & \langle \hat{E}_\alpha^\dag \hat{E}_\beta \rangle_{vqe} & \langle \hat{E}_\alpha^\dag \hat{E}^\dag_\beta \rangle_{vqe}\\
        \langle \hat{E}_\alpha \rangle_{vqe} & \langle \hat{E}_\alpha \hat{E}_\beta \rangle_{vqe} & \langle \hat{E}_\alpha \hat{E}^\dag_\beta \rangle_{vqe}
    \end{bmatrix}\, . \label[equation]{extendedOverlap}
\end{align}

While the idea of extending the subspace is already present in the literature\cite{ollitrault_quantum_2020}, we propose here a new comparison of the approximations obtained in the two subspaces. In addition to a doubling of the number of eigenvalues and eigenstates in this extended subspace, one also expects an increase in precision as a result of having included additional generators and corresponding variational parameters $Y_k$ in the ansatz
\begin{equation}
  \hat{O}_k \ket{\psi_{vqe}} = \left(X_k^0 \hat{I}+ \sum_\alpha (X_k^\alpha \hat{E}_\alpha - Y_k^\alpha \hat{E}_\alpha^\dag) \right)\ket{\psi_{vqe}}\, .
  \label[equation]{qse_extended}
\end{equation}

\begin{figure}[!ht]
  \centering
  \includegraphics[width=\linewidth]{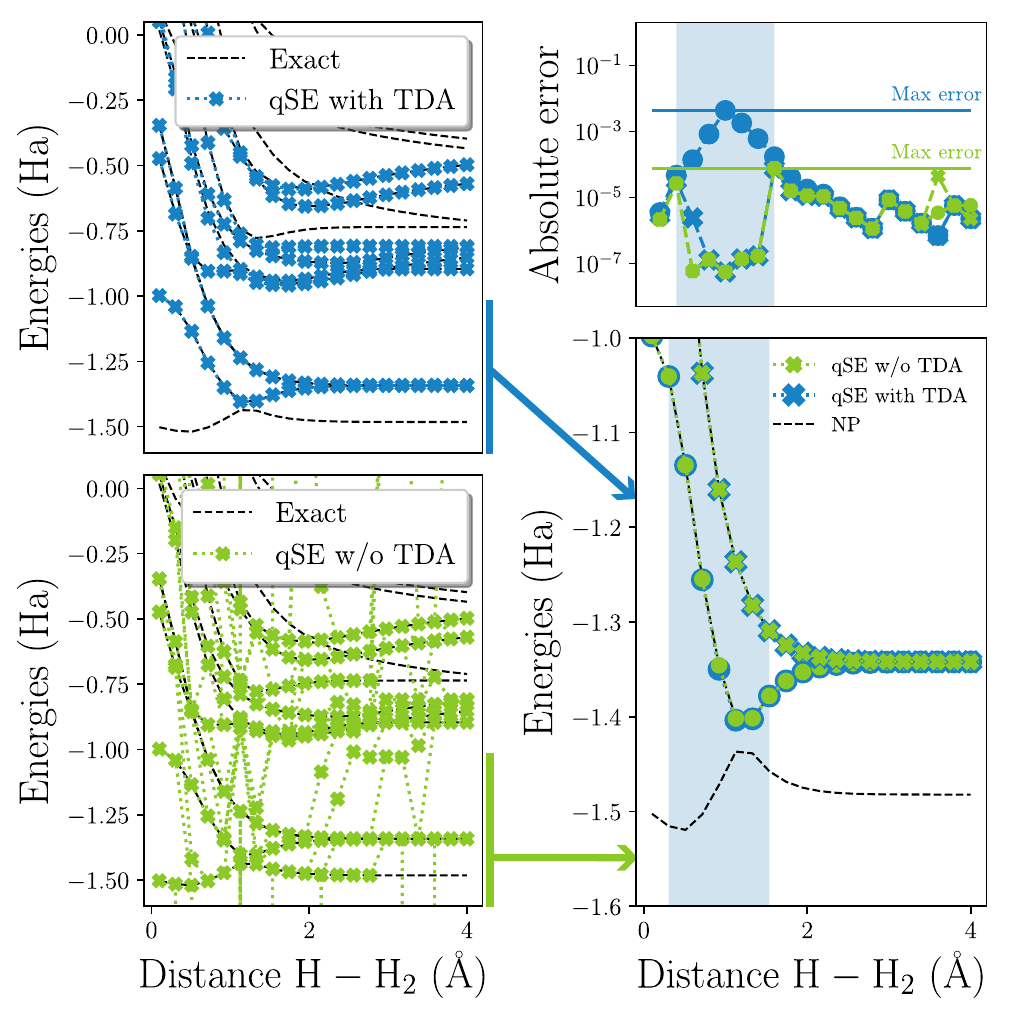}
  \caption{Left panel: Excitation energies obtained for the \ce{H2}-\ce{H} complex for varying distances with the TDA (blue) and beyond the TDA (green). Right panel: Precision of the excitation energies for the first (circles) and second (crosses) excited states calculated with respect to exact diagonalization of the full Hamiltonian. We highlight in light blue the geometries lying in the vicinity of the avoided crossing. In this region, the correlation effects are especially large and consequently, the approximation of the first excited state in $\mathcal{F}_{qse}$ is four orders of magnitude more precise than that in $\mathcal{E}_{qse}$.}
    \label[figure]{ExcitationEnergies}
\end{figure}

In the limit of an uncorrelated ground-state, the de-excitation operators are applied on the mean-field uncorrelated Hartree-Fock state, $\hat{E}^\dag_\alpha\ket{\mathrm{HF}} = 0$. Therefore, they will not contribute to increasing the subspace. In \cref{ExcitationEnergies}, we construct exactly and diagonalize the projected eigenvalue problem in the two subspaces $\mathcal{E}_{qse}$ (with the TDA, blue) and  $\mathcal{F}_{qse}$ (beyond the TDA, green) for the system introduced in \cref{simulationTarget} at different \ce{H}-\ce{H2} distances. We also include exact diagonalization results as reference (black line). Both methods give nearly equivalent results for the second excited state, while including de-excitations operators significantly improves the accuracy of the first excited state in the vicinity of the avoided crossing (green and blue dots in the right panel). This is due to the enhanced expressivity of the chosen subspace with regard to the low-energy spectrum. We also observe numerical instabilities and spurious states when running these calculations (see outliers in bottom left panel). This comes from the ill-conditioning of the $\mathbf{S}_{qse}$ matrix, which indicates the presence of linear dependencies in the set of vectors spanning $\mathcal{F}_{qse}$. These outliers will later become detrimental for molecular dynamics as they require using a matching of the eigenvalues to the corresponding PES instead of a simple ordering.

\subsection{Calculations of excited-state properties and transition amplitudes with subspace-based methods}

In addition to the excitation energies, estimates of other excited-state observables are crucial for many practical applications such as excited-state MD.\cite{Nelson2020_NonAdiabaticDynamics-Review,Park2020_ExcitedStateDynamics-Review} The main observables that we will consider are the excited-state nuclear gradients and the non-adiabatic couplings, which we define and approximate in one spatial dimension $R$ as\cite{Gonzlez2020_ExcitedState-Book}
\begin{equation}
 \begin{aligned}
  F_k(R) &= \frac{\partial \bra{k} \hat{H} \ket{k}}{\partial R}
    \approx \bra{\psi_{vqe}}O_k^\dag \partial_R \hat{H} O_k\ket{\psi_{vqe}} \, , \\
  d_{kl}(R) &= \bra{k} \frac{\partial}{\partial R}\ket{l} 
    \approx \frac{\bra{\psi_{vqe}}O_k^\dag \partial_R \hat{H} O_l\ket{\psi_{vqe}}}{\tilde{\epsilon}_k - \tilde{\epsilon}_l} 
 \end{aligned}
 \label[equation]{nonAdiabaticCouplings}
\end{equation}

To derive the formulas above we assume two approximations. First we neglect the derivatives of the wave function \emph{w.r.t} the nuclear coordinates. Second, we apply the subspace approximation to replace exact eigenstates with approximate ones within the qSE subspace.

The properties introduced in \cref{nonAdiabaticCouplings}, like any second quantized operator $\hat{A}$, can be evaluated by first measuring all the matrix elements of the projected operator
\begin{equation}
    \bm{A}_{qse} =
    \begin{bmatrix}
        \langle \hat{A} \rangle_{vqe} & 
        \langle \hat{A} \hat{E}_\beta \rangle_{vqe} \\
        \langle \hat{E}_\alpha^\dag \hat{A} \rangle_{vqe} & 
        \langle \hat{E}_\alpha^\dag \hat{A} \hat{E}_\beta \rangle_{vqe}
    \end{bmatrix} \, ,
    \label[equation]{Property_Subspace}
\end{equation}
and, later, by classically contracting with the solutions $X_k$ of \cref{qse_pseudoeigenvalue_problem} to yield approximations of excited-state observables and transition amplitudes associated with $\hat{A}$
\begin{equation}
    \label[equation]{qse_matrix_elements}
    \overbrace{\bra{k} \hat{A} \ket{l}}^{\text{exact}} \approx \overbrace{\bra{\psi_{vqe}} O_k^\dag\hat{A} O_l \ket{\psi_{vqe}} =  X_k^\dag \bm{A}_{qse} X_l }^{\text{in}~ \mathcal{E}_{qse}} \, .
\end{equation}

The non-adiabatic couplings in \cref{nonAdiabaticCouplings} are especially challenging to compute with good accuracy in the vicinity of a conical intersection, where the vanishing denominator amplifies the error made in the estimation of the numerator.

\section{Quantum equations of motion: definition and applicability in truncated subspaces}
\label[section]{qeomDefinitionApplicability}

To limit the measurement overhead associated with the estimation of all matrix elements in $\bm{H}_{qse},~\bm{S}_{qse}$, Ollitrault and coworkers proposed an adaptation of the classical equations of motion to the quantum computing setup, namely the quantum Equation of Motion (qEOM) \cite{ollitrault_quantum_2020}. This method relies on the observation that, in the limit of an exact ground state approximation and of an expansion subspace that is sufficiently large to contain the exact excited states, qEOM reduces to standard qSE. In the more realistic case of an approximate ground state and a truncated expansion subspace, only a few studies have been conducted to justify the applicability of the qEOM. We fill this gap by developing in the following section an alternative point of view on the measurement-efficient qEOM in truncated subspaces based on the optimization picture introduced above.

\subsection{qEOM formulation}
Slightly generalising the original definition,\cite{ollitrault_quantum_2020} we refer to the qEOM formulation as a different optimization/eigenvalue problem defined in the expansion subspace of interest (e.g.  $\mathcal{E}_{qse}$). Specifically, one replaces the excitation-energy cost function in \cref{qse_cost_function} with
\begin{equation}
    \label[equation]{qeom_cost_function}
    f_{qeom}(O_k) = \frac{\bra{\psi_{vqe}}[\hat{O}_k^\dag, \hat{H}, \hat{O}_k] \ket{\psi_{vqe}}}
                         {\bra{\psi_{vqe}}[\hat{O}_k^\dag, \hat{O}_k]\ket{\psi_{vqe}}} \, ,
\end{equation}
where the double commutator is defined as 
\begin{align}
    &[A,B,C]= \frac{1}{2}([[A,B],C] + [A, [B,C]])\\
    &= ABC + CBA - \tfrac{1}{2}(ACB+BAC+CAB+BCA)\, . \notag
\end{align}

The form of \cref{qeom_cost_function} has been originally justified by the fact that, assuming that $\ket{\psi_{vqe}}$ is the exact ground state $\ket{0}$ and assuming that the span of the expansion operator includes the exact solutions $\ketbra{k}{0}$ of the eigenvalue problem, then the qEOM and qSE cost functions evaluated at these particular points coincide,\cite{ollitrault_quantum_2020} \ie
\begin{align}
  \label[equation]{qeom_cost_function_reduced}
  f_{qeom}(\vert k \rangle \langle 0 \vert)&= f_{qse}(\vert k \rangle \langle 0 \vert) - \epsilon_0.
\end{align}

Note that the numerator and denominator in \cref{qeom_cost_function} are also bilinear in the trial operators, and one can expand the trial operator as a linear combination of the intermediary states. Therefore, the extrema under the orthogonality conditions
\begin{equation}
    \label[equation]{qeom_orthogonality_constraints}
    \bra{\psi_{vqe}}[\hat{O}_k^\dag, \hat{O}_i]\ket{\psi_{vqe}} =
    X_k^\dag \bm{S}_{qeom} X_i = \delta_{ik} \, .
\end{equation}
are the solutions of a pseudoeigenvalue equation of the form
\begin{equation}
    \label[equation]{qeom_pseudoeigenvalue_problem}
    \bm{H}_{qeom} X_k = \bm{S}_{qeom} X_k \tilde{\epsilon}_{k0}^{qeom}\, .
\end{equation}

For general fermionic expansion in $\mathcal{E}_{qse}$, we introduced the linear ansatz state $\hat{O}_k\ket{\psi_{vqe}} = X_k^0 \hat{I} \ket{\psi_{vqe}} + \sum_\alpha X_k^\alpha \hat{E}_\alpha \ket{\psi_{vqe}}$ and the corresponding ansatz operator $\hat{O}_k = X_k^0 \hat{I} + \sum_\alpha X_k^\alpha \hat{E}_\alpha$. From the properties of the commutators, it follows that the cost function in \cref{qeom_cost_function} is invariant upon the transformation $\hat{O}_k \rightarrow \hat{O}_k + a \hat{I}$ for all values of $a$. The choice of $a = - \bra{\psi_{vqe}}\hat{O}_k \ket{\psi_{vqe}}$, in particular, suggests a more appropriate expansion of the trial operators in the qEOM problem 
\begin{equation}
\label[equation]{qeom_ansatz}
    \hat{O}_k = \sum_\alpha X_k^\alpha \left(\hat{E}_\alpha - \bra{\psi_{vqe}} \hat{E}_\alpha \ket{\psi_{vqe}}\hat{I}\right)\, .
\end{equation}
Introducing $:\mathrel{\hat{E}_\alpha}: = \hat{E}_\alpha - \bra{\psi_{vqe}} \hat{E}_\alpha \ket{\psi_{vqe}}\hat{I}$ the translated generators,
and using the commmutator invariance to substituting the shifted $:\mathrel{\hat{E}_\alpha}:$ with the generators $\hat{E}_\alpha$, we find the following projected operators for the qEOM problem,
\begin{align}
    \bm{H}_{qeom}& =
    \begin{bmatrix}
        \langle [\hat{E}_\alpha^\dag, \hat{H}, \hat{E}_\beta] \rangle_{vqe}
    \end{bmatrix}, \label{H_qeom} \\
    \bm{S}_{qeom}& =
    \begin{bmatrix}
        \langle [\hat{E}_\alpha^\dag, \hat{E}_\beta] \rangle_{vqe}
    \end{bmatrix} \, . \label{S_qeom}
\end{align}

In the qSE problem we expected that an expansion subspace including the identity in its generators will necessarily contain a more accurate approximation of the ground state. Instead, a key consequence of the previous invariance of the qEOM problem is that it is only formulated in the subspace orthogonal to the initial state $\ket{\psi_{vqe}}$ by virtue of the gauge fixing condition and thus cannot be used for improving or mitigating the ground state approximation.

A notable advantage of the qEOM formulation is that the fermionic rank of the observables to be calculated to construct the projected Hamiltonian is greatly reduced compared to qSE. While qSE requires the evaluation of the two- and three-particle Reduced Density Matrices (RDMs), qEOM only requires one- and two-particle RDMs because each commutator reduces the fermionic rank by one compared to the corresponding operator product, see \cref{Appendix_2}. 

However, the equality \cref{qeom_cost_function_reduced} of the two cost functions at the exact solutions does not generalize to the corresponding truncated eigenvalue problem when the expansion subspace does not contain these solutions. Therefore, in the general case, the qEOM and qSE pseudo-eigenvalue problems are not related and the eigenvalues of the first are not necessarily good approximations of the qSE excitation energies.

\subsection{\texorpdfstring{qEOM formulation in $\mathcal{F}_{qse}$}{qEOM formulation in Fqse}}
With the same argument used to derive the qEOM pseudo-eigenvalue problem in  $\mathcal{E}_{qse}$, we can extend the definition of the qEOM projected operators in $\mathcal{F}_{qse}$
\begin{align}
    \bm{H}_{qeom} & =
    \begin{bmatrix}
        \langle [\hat{E}_\alpha^\dag, \hat{H}, \hat{E}_\beta] \rangle_{vqe} & \langle [\hat{E}_\alpha^\dag, \hat{H}, \hat{E}^\dag_\beta] \rangle_{vqe}\\
        \langle [\hat{E}_\alpha, \hat{H}, \hat{E}_\beta] \rangle_{vqe} & \langle [\hat{E}_\alpha, \hat{H}, \hat{E}^\dag_\beta] \rangle_{vqe}
    \end{bmatrix}, \label[equation]{extendedHamiltonianF}\\
    \bm{S}_{qeom} & =
    \begin{bmatrix}
        \langle [\hat{E}_\alpha^\dag, \hat{E}_\beta] \rangle_{vqe} & \langle [\hat{E}_\alpha^\dag , \hat{E}^\dag_\beta] \rangle_{vqe}\\
        \langle [\hat{E}_\alpha, \hat{E}_\beta] \rangle_{vqe} & \langle [\hat{E}_\alpha, \hat{E}^\dag_\beta] \rangle_{vqe}
    \end{bmatrix} \, . \label[equation]{extendedOverlapF}
\end{align} 

Remarkably, the projected qSE matrices have entire vanishing blocks in the weak-correlation limit, while the diagonal blocks of the projected qEOM matrices are non-vanishing because of the use of commutators. For this reason, we observed in our calculations numerical instabilities when diagonalizing the qSE pseudo-eigenvalue problem in $\mathcal{F}_{qse}$, which did not happen for the qEOM pseudo-eigenvalue problem.

Furthermore, we also remark here that, even if this does not justify the validity of qEOM in $\mathcal{F}_{qse}$, the qEOM formulation can be regarded as extracting the antisymmetric part of the qSE problem. The underlying anti-symmetry takes the form $\hat{O}_k \leftrightarrow \hat{O}_k^\dag$, or equivalently expressed as $\hat{E}_\alpha \leftrightarrow \hat{E}_\alpha^\dag$, which is reminiscent of particle-hole symmetry.
This symmetry implies that the solutions of the qEOM pseudo-eigenvalue problem appear in pairs ($\tilde{\epsilon}_{k0}^{qeom}, -\tilde{\epsilon}_{k0}^{qeom}$) and also reduces the number of measurements required to build the projected operators. We found that these elements also contributed to the numerical stability of the approach and discuss them further in \cref{Appendix_2}.

\subsection{General considerations of the qEOM formulation in truncated subspaces}
We detail in the following the role of each assumption in the derivation of the qSE/qEOM equality (cf. \cref{qeom_cost_function_reduced}) at the exact solution points, with the aim of more generally relating the qEOM and qSE cost functions \emph{for any} trial operator $\hat{O}_k$ in the expansion subspace.

Under the first assumption $\vert \psi_{vqe} \rangle = \vert 0 \rangle$ only, the form of the qEOM cost function can be reduced for any trial operator in the subspace to 
\begin{equation}
  \label[equation]{qeom_exact_groundstate}
  f_{qeom}(\hat{O}_k) =  \frac{\bra{0} \hat{O}_k^\dag \Delta \hat{H} \hat{O}_k + \hat{O}_k \Delta \hat{H} \hat{O}_k^\dag \ket{0}}
                        {\bra{0} \hat{O}_k^\dag \hat{O}_k - \hat{O}_k \hat{O}_k^\dag \ket{0}} \, ,
\end{equation}
where $\Delta \hat{H} = \hat{H} - \epsilon_0$ is the Hamiltonian shifted \emph{w.r.t} the exact ground state energy. This form highlights the relation with the initial qSE cost function. We provide in \cref{noExactGroundState} a similar expression when the exact ground state assumption is relaxed, but we note that the following derivations do not hold in the general case. $\Delta\hat{H}$ and $\hat{I}$ are respectively positive semi-definite and positive definite such that, $\bra{0} \hat{O}_k \Delta \hat{H} \hat{O}_k^\dag \ket{0}$, $\bra{0} \hat{O}_k^\dag \Delta \hat{H} \hat{O}_k \ket{0}$ and $\bra{0} \hat{O}_k \hat{O}_k^\dag \ket{0}$, $\bra{0} \hat{O}_k^\dag \hat{O}_k \ket{0}$ are positive valued. This first observation guarantees the ordering 
\begin{equation}
\label[equation]{qeom_qse_cost_function_inequality}
  f_{qeom}(\hat{O}_k) \geq f_{qse}(\hat{O}_k) - \epsilon_0 \, ,
\end{equation} 
assuming that $\|\hat{O}_k^\dag \ket{0}\|^2 \leq \| \hat{O}_k \ket{0} \|^2$ holds. Because this condition is always satisfied by one element of the pair ($\hat{O}_k$, $\hat{O}_k^\dag$), we can conclude that the positive values of the qEOM cost function are lower bounded by the minimum of the qSE cost function which, by definition, is the optimal solution in the variational subspace. This lower bound is already a strong guarantee that the qEOM eigenvalues are well-motivated proxies for approximating the Hamiltonian's eigenvalues without evaluating the qSE eigenvalues.

Under the second assumption that the subspace contains the exact excited states, we find that the minimum of the qSE and qEOM cost functions coincide and the lower bound \cref{qeom_qse_cost_function_inequality} becomes the strict equality \cref{qeom_cost_function_reduced} at these points. In a truncated expansion subspace, this assumption will not necessarily hold and $f_{qeom}$ and $f_{qse}$ typically have different extrema that both differ from the exact solutions. More importantly, the non-standard form of the orthogonality conditions in \cref{qeom_orthogonality_constraints} or equivalently, of the overlap matrix $\bm{S}_{qeom}$ in \cref{S_qeom} only reduces to the usual orthogonality of the corresponding states for the exact solutions in the full Hilbert space. Thus, the excited states  $\ket{\psi_{qeom}^k} = \hat{O}_k \ket{\psi_{vqe}}$ constructed as solutions of the qEOM pseudo-eigenvalue problem in a truncated subspace will not \emph{a priori} be mutually orthogonal. The condition for maintaining the desired orthogonality of the approximated excited states is captured by the so--called vacuum condition (VAC),\cite{prasad_aspects_1985}
\begin{equation}
  \hat{O}_k^\dag \ket{\psi_{vqe}} = 0 \, ,
  \label[equation]{VAC}
\end{equation}
that must be satisfied by the solutions of the pseudo-eigenvalue problem, or equivalently, of the constrained optimization problem.

 \subsection{Vacuum condition}
 
Despite the computational interest of the formulation of qEOM in terms of commutators defined in \cref{qeom_cost_function}, the loss of the excited states orthogonality due to the possible violation of the VAC was an issue already identified for classical methods based on the equations of motion.\cite{rowe_equations_of_motion_1968, prasad_aspects_1985, szekeres_killer_2001}
This issue can be circumvented in the case of $\mathcal{E}_{qse}$, by introducing an alternative set of so--called \emph{self-consistent} expansion operators $\mathcal{S}_{qse} = \{\hat{S}_\alpha, ~\forall \alpha\}$ defined from a similarity transformation of the conventional fermionic excitation operators $\hat{E}_\alpha$\cite{Kim2023_Davidson-scQEOM,asthana_quantum_2023}, such that
\begin{equation}
   \hat{S}_\alpha^\dag \ket{\psi_{vqe}} = 0 \, , \quad  \forall \alpha \, .
  \label{eq:VacuumCondition}
\end{equation}

Under the additional assumption that the approximated ground state is exact, this automatically ensures that the VAC is fulfilled for any trial state, linear combination of the self-consistent expansion operators and in particular for the solutions of the pseudo-eigenvalue problem in the truncated subspace.

Our goal is, however, different: we do not want to impose \emph{a priori} the VAC at the level of the expansion operators, which corresponds to picking an expansion subspace in which the commutator and product are formally equivalent and do not lead to a reduced measurement overhead.
We instead aim at critically comparing the accuracy of the qSE and qEOM approaches based on the same fermionic expansion subspace (\eg  $\mathcal{E}_{qse}$ or  $\mathcal{F}_{qse}$).
Specifically, we aim to investigate how the loss in formal orthogonality between the excited states, which is the price to pay for the reduced scaling of qEOM, affects the accuracy of excited-state properties.

\subsection{Applicability of qEOM}

To have any reduction of the measurement cost from using the qEOM formulation instead of the qSE equivalent, we argue that the VAC should not be strictly imposed for any operators of the subspace, \ie at the level of the generators. Instead, we understand the qEOM cost function as the joint minimization of the desired excitation energy and of a quadratic penalty term for trial operators that do not satisfy the VAC. If good approximations of the exact solutions are contained in the expansion subspace, then the penalty term will approach zero at the minima of the qEOM cost function.

\subsubsection{\texorpdfstring{Applicability in $\mathcal{E}_{qse}$}{Applicability in Eqse}}

In $\mathcal{E}_{qse}$, we find that one can actually provide bounds for this penalty term based on a very natural hypothesis relating the true ground state $\ket{0}$ and the mean-field Hartree-Fock state $\ket{\mathrm{HF}}$ 
\begin{equation} \label[equation]{exact_groundstate_hf}
    \ket{0} = \sqrt{1-\gamma^2}\ket{\mathrm{HF}} + \gamma \ket{u} \, .
\end{equation}

We introduced here the unitary vector $\ket{u}$, orthogonal to the Hartree-Fock state, and a parameter $\gamma$. Small $\gamma$ values correspond to the weakly correlated regime for which the fermionic expansion subspace was initially introduced. This equation yields for the adjoint of the expansion operators $E_\alpha^\dag \ket{0} = \gamma E_\alpha^\dag \ket{u}$ as the Hartree-Fock state annihilates all the fermionic deexcitations $E_\alpha^\dag$, which, at zeroth order in $\gamma$, is exactly the definition of the self-consistent operators. However, we can additionally derive the penalty term at the second order in $\gamma$
\begin{align}
  \label[equation]{qeom_second_order}
  &\frac{f_{qeom}(\hat{O}_k)}{f_{qse}(\hat{O}_k)} = \notag\\
  & 1 + \gamma^2\left(\frac{\bra{u} \hat{O}_k \Delta \hat{H} \hat{O}_k^\dag \ket{u}}{\bra{0} \hat{O}_k^\dag \Delta \hat{H} \hat{O}_k \ket{0}} + \frac{\bra{u} \hat{O}_k \hat{O}_k^\dag \ket{u}}{\bra{0} \hat{O}_k^\dag \hat{O}_k \ket{0}}\right) + \mathcal{O}(\gamma^4)\, .
\end{align}

Remarking that $\|\hat{O}_k^\dag \ket{u}\|^2 \leq \|\hat{O}_k^\dag \ket{0}\|^2 \leq \| \hat{O}_k \ket{0} \|^2$, where the first part is a triangular identity and the second was assumed in deriving \cref{qeom_qse_cost_function_inequality}, the factor associated with $\gamma^2$ is upper bounded by a constant that we don't try to identify in this work.

In addition, the orthogonality constraints for the optimization of the excited states are given by
\begin{equation}
    \bra{0}[\hat{O}_k^\dag, \hat{O}_j]\ket{0} = \bra{0}\hat{O}_k^\dag \hat{O}_j\ket{0} - \gamma^2 \bra{u}\hat{O}_j \hat{O}_k^\dag \ket{u} \, .
\end{equation}

Again, the fact that $\gamma$ appears at the second order in these formulas is a strong indication that, even in the worst case, the eigenvalues of the qEOM pseudo-eigenvalue problem in $\mathcal{E}_{qse}$ will be good approximations of the qSE eigenvalues.

\subsubsection{\texorpdfstring{Applicability in $\mathcal{F}_{qse}$}{Applicability in Fqse}}
The above guarantees that the qEOM formulation, in $\mathcal{E}_{qse}$ and in the weak-correlation limit, always produces reliable estimates of the excited states cannot be extended to $\mathcal{F}_{qse}$. This is because we explicitly introduced $\mathcal{F}_{qse}$ to account for large correlation effects, while the guarantees we proposed depend on worst-case bounds for the additional terms appearing in $f_{qeom}$ in the weak-correlation limit. Instead in $\mathcal{F}_{qse}$, we can only rely on the assumption that the fermionic subspace is a good heuristic for the low-energy spectrum of the system. We also remark here that the loss of orthogonality identified for the qEOM excited states, has, by construction, a reduced effect on the low-energy part of the spectrum. Indeed, the lowest-lying excited states are defined by fewer orthogonality constraints than the excited states of greater energy, as can be seen in \cref{qeom_orthogonality_constraints}.

For the low-energy spectrum of the target system in this work, we already identified in \cref{ExcitationEnergies} that the limited expressivity of $\mathcal{E}_{qse}$ in the vicinity of the conical intersection could be overcome by adding de-excitations in the set of generators and considering the extended subspace $\mathcal{F}_{qse}$ instead. We thus do not expect the expressivity of the fermionic subspace to be a limiting factor for the qEOM excited-state approximations.

\subsection{qEOM for excited state properties}
Once the qSE (or qEOM) eigenvalue problem has been solved, one can estimate additional properties of the excited states with extra measurements and using the formulas in \cref{nonAdiabaticCouplings}. Evaluating a general observable $\hat{A}$ that is not diagonal in the excited state basis requires calculating the products of the generators and the target properties instead of the commutators. For this reason, the reduction of the number of measurements characterizing the qEOM formulation does not apply to the calculation of general excited-state properties from the qEOM eigenvectors.

In addition, the gauge freedom, which was implicitly fixed in the calculation of excitation energies, needs to be explicitly enforced in the calculation of any non-diagonal properties. To compute these properties, we first explicitly build the translated generators $:\mathrel{\hat{E}_\alpha}:$ and, thus, we measure all ground-state like expectation values $\bra{\psi_{vqe}}\hat{E}_\alpha \ket{\psi_{vqe}}$. The number and size of these measurements remain small and does not affect, in terms of scaling, the reduction of the measurement overhead in the qEOM formulation. We can then estimate the matrix elements $\langle \hat{A} \rangle_{vqe}$, $ \langle \mathrel{\hat{A}} :\mathrel{\hat{E}_\beta}: \rangle_{vqe}$, $\langle :\mathrel{\hat{E}_\alpha^\dag}: \mathrel{\hat{A}} \rangle_{vqe}$, $ \langle :\mathrel{\hat{E}_\alpha^\dag}: \mathrel{\hat{A}} :\mathrel{\hat{E}_\beta}: \rangle_{vqe}$ with the translated excitation operators and build the projected operators $\mathbf{A}_{qse}$.

As an example, consider the estimation of excitation energies with the qEOM approach. Because of potential non-vanishing penalty terms, the qEOM eigenvalues $\tilde{\epsilon}_{k0}^{qeom} = f_{qeom}(O_k)$ do not exactly coincide with the energy of the states $O_k\ket{\psi_{vqe}}$, which can be calculated as excited-state properties and which we call the \emph{unpenalized excitation energies}
\begin{equation}
  \label[equation]{unpenalized_energy_estimators}
  \tilde{\epsilon}_{k0}^{w/o~penalty} = f_{qse}(O_k) = \frac{\bra{\psi_{vqe}} \hat{O}_k^\dag \Delta \hat{H} \hat{O}_k \ket{\psi_{vqe}}}
  {\bra{\psi_{vqe}} \hat{O}_k^\dag \hat{O}_k \ket{\psi_{vqe}}} \, .
\end{equation}

\begin{figure}[!ht]
  \includegraphics[width=\linewidth]{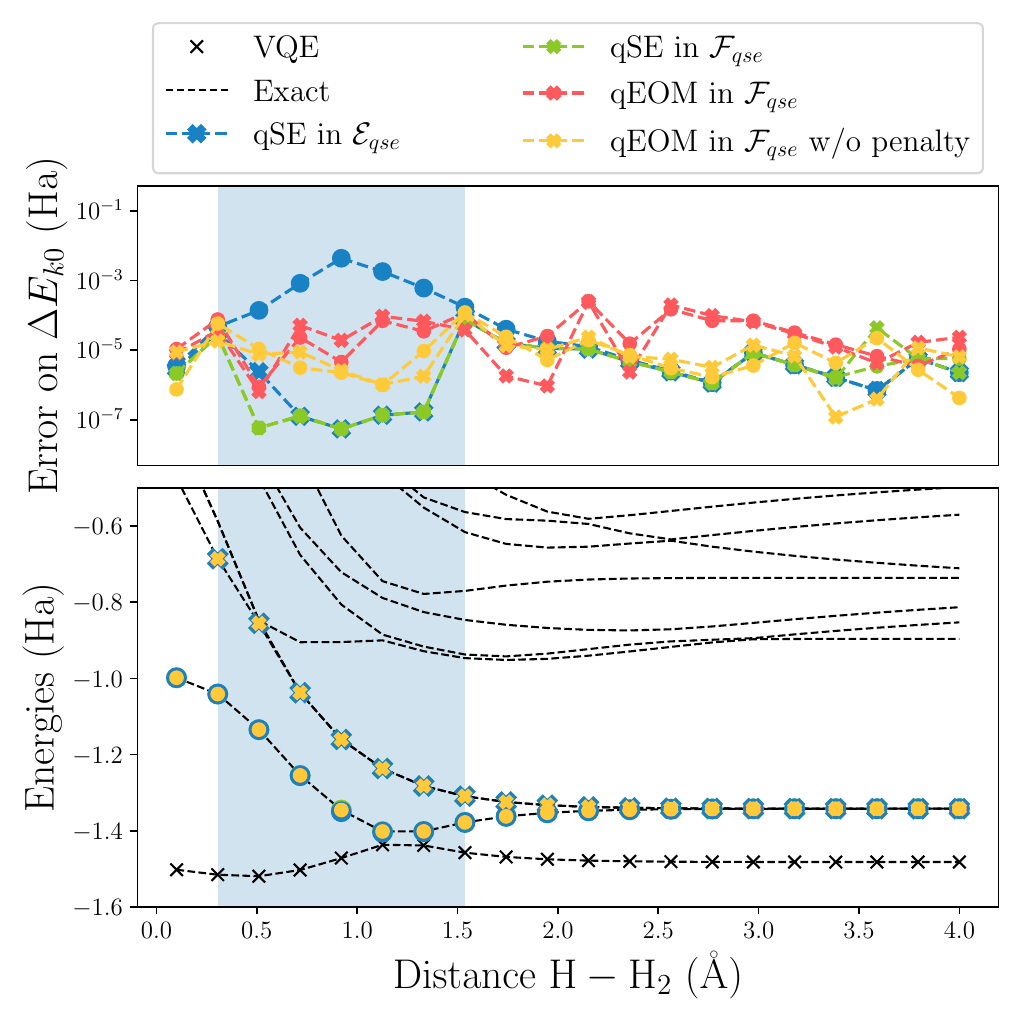}
  \caption{
  Accuracy of different flavors of subspace expansion methods in $\mathcal{F}_{qse}$, with respect to the excitation energies  calculated with exact diagonalization in the full Hilbert space (black lines). We also represent (in blue) the initial results obtained with the qSE approach in the TDA, reference upon which we aim to improve.
  We report in the upper panel the error on the excitation energies of the first (circles) and second (crosses) excited states.
  We report in the lower panel the corresponding excitation energies for varying \ce{H2}-\ce{H} distance.
  The expensive qSE method without the TDA (green) yields the best energy approximations in the extended subspace. The excitation energies obtained from the qEOM pseudo-eigenvalues (red) also exhibit the improved accuracy characterizing $\mathcal{F}_{qse}$ close to the avoided crossing (highlighted in light blue). However, these eigenvalues still include small contributions from the penalty terms, as can be seen calculating the unpenalized excitation energies (gold), which show a better agreement with the best approximations in the subspace.}
  \label[figure]{energy_comparison_qse_qeom}
\end{figure}

In \cref{energy_comparison_qse_qeom}, we compare the precision of the first two excited states (circles and crosses) for different flavors of subspace expansion in $\mathcal{F}_{qse}$ with respect to two references: the exact diagonalization in the full Hilbert space (black dashed lines) and the qSE estimates in the smaller subspace \ie in the TDA (blue). As already seen in \cref{ExcitationEnergies}, the inclusion of the de-excitations beyond the TDA (green) improves the accuracy of the first excited states in the vicinity of the avoided crossing. This improvement is also present when estimating the excitation energies from the cost-efficient eigenvalues of the qEOM problem (red), despite an overall loss of accuracy of up to three orders of magnitude over the qSE solutions in $\mathcal{F}_{qse}$. This is partly due to the penalty term accounted for in the qEOM eigenvalues, as can be seen when calculating the unpenalized energy of the qEOM optimized states $\hat{O}_k \ket{\psi_{vqe}}$ (yellow). While a small difference remains close to the avoided crossing, these reconstructed excitation energies show good agreement with the best solutions in the subspace along the whole dissociation range.

\section{Applications to non-adiabatic molecular dynamics with quantum computing}
\label[section]{applicationToMolecularDynamics}

The accuracy of the unpenalized qEOM energy estimates is a good indication that the excited states properties, which are calculated from the eigenvectors of the qEOM pseudo-eigenvalue problem and not from its eigenvalues, will exhibit the improved accuracy expected from the extended subspace. With the absence of outliers in the spectrum and the reduction of the measurement overhead for the construction of the pseudo-eigenvalue problem, these are key practical advantages of the qEOM formulation for our application. 





\subsection{Comparison of qSE and qEOM on properties influencing non-adiabatic molecular dynamics}

\begin{figure*}
    \centering
    \includegraphics[width=\linewidth]{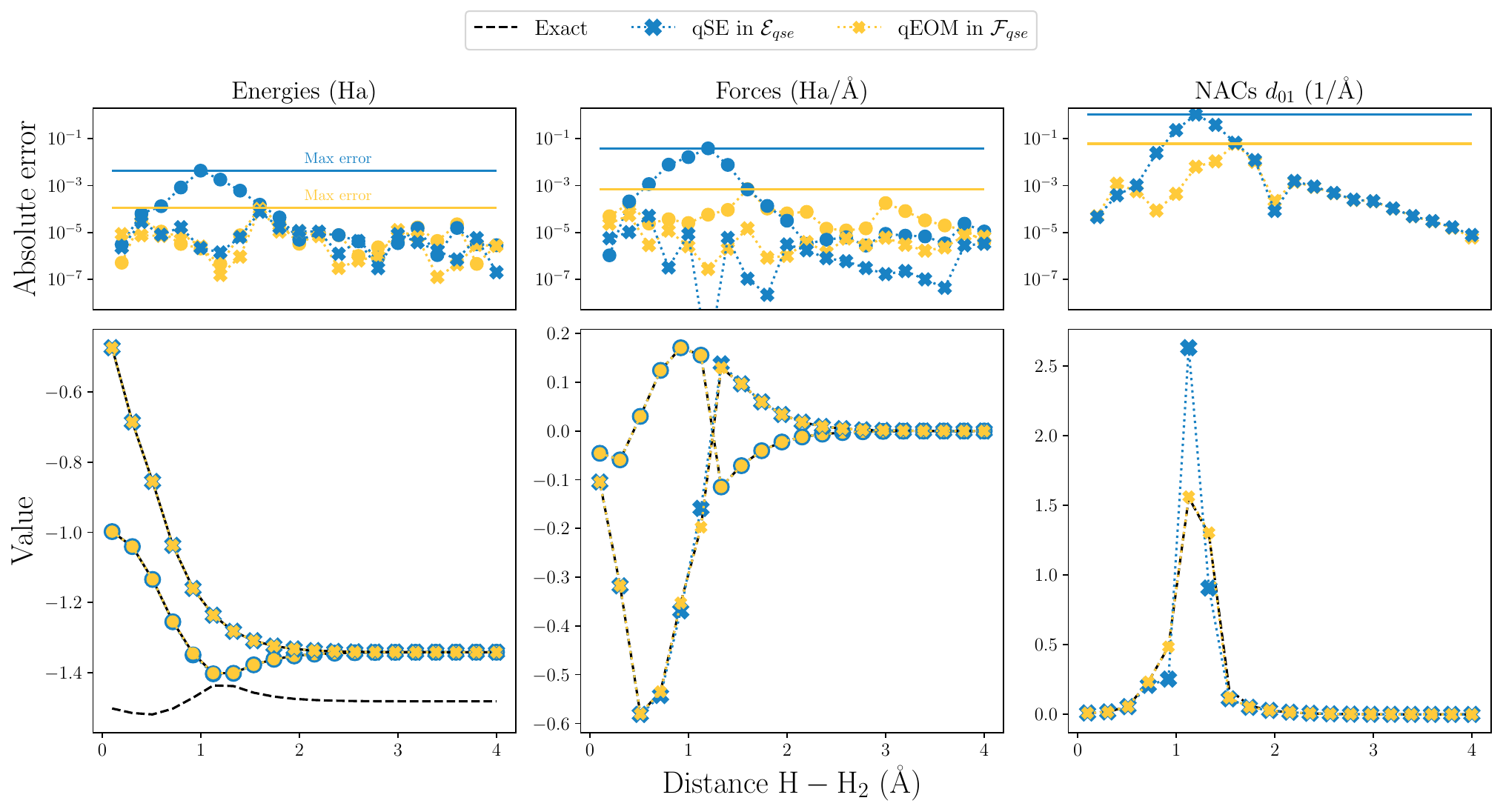}
    \caption{
    Comparison of qSE in the TDA (blue) and qEOM (yellow) for the estimation of the first (circles) and second (crosses) excitation energies (left), adiabatic nuclear energy gradients (middle) and the non-adiabatic couplings $d_{01}$ (right). Note that the non-adiabatic couplings in the third panel are associated with two eigenstates, but we still represent them with crosses given that, in this work, we only considered couplings with the ground state $d_{0k}$.
    The horizontal lines in the error subplots represent the maximum error with either techniques, which is systematically obtained in the region of the conical intersection. We do not include here the results of the qSE approach in the extended subspace which also feature an improved precision close to the conical intersection.}
    \label[figure]{Energies_Forces_Nacs}
\end{figure*}

We now examine the accuracy of two methods described above on the three properties relevant to molecular dynamics: excitation energies, adiabatic nuclear energy gradients and non-adiabatic couplings.
As a reference, we include the qSE results in the TDA. We also include the qEOM results expressed in the enlarged subspace $\mathcal{F}_{qse}$. Note that we could have included the qSE results in the extended subspace and that they would also have featured an improved accuracy close to the conical intersection. However, the relevant observable for calculations of molecular dynamics, the derivative of the Hamiltonian, is a single-body observable. We can thus avoid calculating the 3-particle Reduced Density Matrices (RDMs) if we consider the qEOM approach instead of the qSE one (see \cref{Appendix_2}).

In~\cref{Energies_Forces_Nacs}, we compare the accuracy of these two methods for all three properties over the full dissociation path. At the vicinity of the conical intersection between the ground state and the first excited state, correlation effects become larger and, consequently, the TDA results become less accurate with absolute errors up to $10^{-2}~\mathrm{Ha}$ 
for the energies,  $10^{-2}~\mathrm{Ha}~\textrm{\AA}^{-1}$ for the forces,
and up to $10^{0}~\textrm{\AA}^{-1}$ for the non-adiabatic couplings.
This indicates that even a small inaccuracy in the excited-state wave function is amplified in the vicinity of the avoided crossing.
Far from the crossing, we observe a good agreement between both methods for the two excited states.

For the system investigated here, we find that including the de-eexcitations in the generators of the expansion subspace beyond the TDA leads to an improved global accuracy for all three observables relevant to molecular dynamics, due to the significantly more precise description of the region with larger correlation effects. This improved accuracy also applies to the qEOM formulation of the problem, despite the loss of strict mutual orthogonality of the approximate excited states.


\subsection{Simulation of the non-adiabatic population transfer with qSE and qEOM}

Based on the properties of the lowest-lying excited-states computed above, we run molecular dynamic simulations using Tully's fewest switches\cite{Tully1990_FSSH} algorithm.
We initialize a ground state wavepacket by sampling initial conditions, position and velocity, for a swarm of virtual particles which we individually evolve in time.
The PESs and the adiabatic forces allow simulating the adiabatic time evolution of the nuclei within the Born-Oppenheimer approximation.
The amplitude of the non-adiabatic couplings determines the transition probability of the virtual particles between two PESs.
In this way, non-adiabatic transitions are included on average through the stochastic evolution of each virtual particle.
Note that we do not consider exact vibrational dynamics simulations and, therefore, we do not discuss the limitations of the Tully's Fewest Switches approach for molecular dynamics.

\begin{figure}
  \centering
  \includegraphics[width=\linewidth]{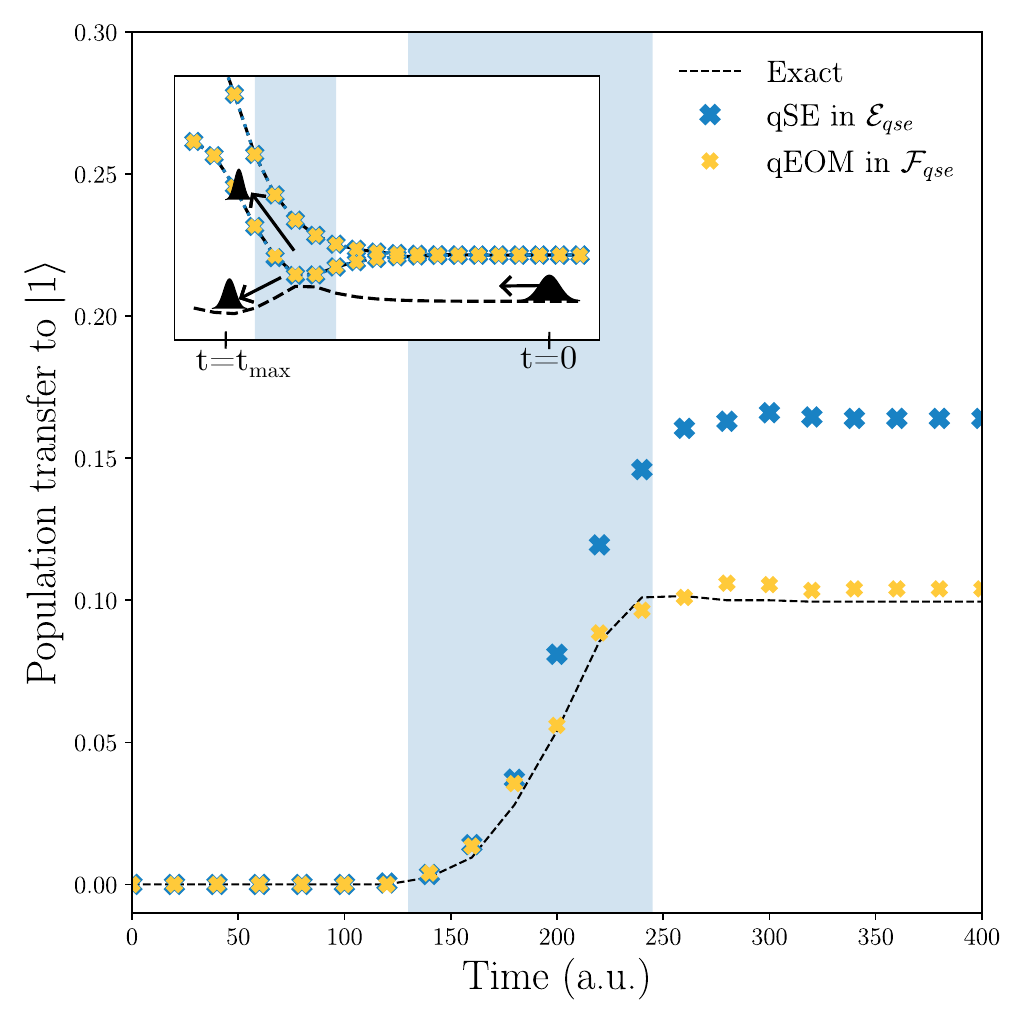}
  \caption{
  Population transfer calculated running Tully's Fewest Switches molecular dynamic simulations with a swarm of particles representing the ground state wavepacket and crossing the conical intersection around $t\approx 200 s$.
  We interpolate the values of the energies, forces, and non-adiabatic couplings calculated with the qEOM approach (yellow crosses), the qSE approach in the TDA (blue crosses) and with the exact diagonalization (black line) and simulate the dynamics on the interpolated surfaces. We highlight in blue the region in which the non-adiabatic transfer takes place.}
  \label[figure]{PopTransfer}
\end{figure}

Converging the calculation of the population transfer probability requires especially accurate calculations of the energies, forces, and couplings over the full dissociation range.
Because of the error amplification in the calculation of the non-adiabatic couplings, we found that the evaluation of the couplings is key to achieve quantitatively converged simulations.
In \cref{PopTransfer} we report the final population of the first excited state based on data calculated with the qSE approach in the TDA and with the qEOM approach beyond the TDA.
As observed above, the error on the properties calculated with the former method are large and, therefore, the population transfer deviates qualitatively from the exact value.
For qEOM, we instead observe a good qualitative agreement which is a direct consequence of the improvement of the improved energy, force, coupling estimates in the region of the conical intersection. 

\section{Conclusions}
\label[conclusion]{conclusions}

We presented a fully-integrated theoretical and computational framework for non-adiabatic molecular dynamics simulations within the fewest-switches surface-hopping strategy, together with its implementation into the quantum computing software package Qiskit\cite{Qiskit}. 
The proposed developments allow for the calculation of the excited-state properties required to simulate non-adiabatic dynamics -- i.e., the potential energy surfaces, their gradients, and the non-adiabatic couplings between them -- with different flavors of quantum subspace expansion and quantum equation of motion techniques.
In this respect, we discussed how qualitatively accurate molecular dynamic calculations of a seemingly simple test case, namely the \ce{H2}-\ce{H} collision process, already require additional tailoring over the standard subspace expansion technique. Specifically, we show how the qEOM improves the precision of these calculations compared to  qSE methods based on the TDA and avoids unphysical pseudo-eigenvalues rendering impractical the quantum subspace expansion approaches. 
Up to now, the nuclear quantum dynamics is described within the semiclassical surface-hopping method.
However, the framework can be straightforwardly interfaced with fully quantum vibrational dynamics methods such as, for instance, the multi-configurational time-dependent Hartree method~\cite{Worth2008_MCTDH-Review}. Alternatively, it can be combined with a multi-component electron-nuclear quantum dynamics such as Nuclear-Electronic Orbital (NEO) framework~\cite{webb2002multiconfigurational}.
In the first case, one can possibly leverage quantum hardware to speed up the vibrational dynamics itself~\cite{Miessen2022_QD-QC-Review,Ollitrault2022_GridBased-QD}, while efficient quantum computing implementations of the NEO algorithms have already shown promising results~\cite{kovyrshin2023quantum,kovyrshin2023quantum_dyn}.

Future extensions will include the scaling-up towards larger and potentially more complex molecular systems.
We plan to combine the approaches presented in this paper with the embedding schemes recently developed in our team~\cite{Rossmannek2023_ProjectionBasedEmbedding}.
Additionally, we expect that the same approach presented in this work can also be extended to other excited-state properties, such as spin-orbit couplings~\cite{Marian2012_SpinOrbitCoupling} or high-order response properties~\cite{Helgaker2012_Review}.

Finally, a thorough analysis of the robustness of the proposed algorithms and, in particular, of subspace expansion techniques, against circuit noise in realistic quantum computing settings is required before proceeding with near-term quantum hardware calculations. We plan to address this issue in future investigations. 

\section{Acknowledgments}
The authors thank Max Rossmannek, Jakob Braun, Laurin Fischer for useful discussions.
IBM, the IBM logo, and ibm.com are trademarks of International Business Machines Corp., registered in many jurisdictions worldwide. Other product and service names might be trademarks of IBM or other companies. The current list of IBM trademarks is available at https://www.ibm. com/legal/copytrade.
Work on ``Quantum Computing for Photon-Drug Interactions in Cancer Prevention and Treatment'' is supported by Wellcome Leap as part of the Q4Bio Program.

\nocite{*}
\bibliography{bibliography}

\newpage
\appendix

\section{qEOM calculations without assuming an exact ground state}
\label[section]{noExactGroundState}
We consider here the general situation where the expansion subspace is built from an optimized state which can be expressed as 
\begin{equation} \label[equation]{inexact_groundstate}
    \ket{\psi_{vqe}} = \sqrt{1-\beta^2}\ket{0} + \beta \ket{v} \, ,
\end{equation}
with a parameter $\beta$ controlling the quality of the ground state approximation, and where $\ket{v}$ is orthogonal to the ground state and quantifies the deviation of $\ket{\psi_{vqe}}$ from $\ket{0}$. We also introduce the energy of the optimized VQE state $\epsilon_{vqe} = \bra{\psi_{vqe}} \hat{H} \ket{\psi_{vqe}} = (1-\beta^2)\epsilon_0 + \beta^2\bra{v}\hat{H}\ket{v}$.

Contrary to the previous case, the reference state $\ket{\psi_{vqe}}$ is no longer an exact eigenstate of the Hamiltonian and the double commutator cannot be simplified. We introduce the notation $\mathcal{L}(A,B,C) = ABC - \frac{1}{2}(ACB + BAC)$ and derive an equation analogous to \cref{qeom_exact_groundstate}

\begin{equation}
  \label[equation]{qeom_inexact_groundstate}
  f_{qeom}(\hat{O}_k) =  \frac{\langle \mathcal{L}(\hat{O}_k^\dag, \hat{H}, \hat{O}_k) + \mathcal{L}(\hat{O}_k, \hat{H}, \hat{O}_k^\dag) \rangle_{vqe}}{\langle \hat{O}_k^\dag \hat{O}_k - \hat{O}_k \hat{O}_k^\dag \rangle_{vqe}} \, ,
\end{equation}
where 
\begin{equation}
    \langle \mathcal{L}(\hat{O}_k^\dag, \hat{H}, \hat{O}_k)\rangle_{vqe} = \langle \hat{O}_k^\dag \hat{H} \hat{O}_k - \hat{H}\hat{O}_k^\dag \hat{O}_k - \hat{O}_k^\dag\hat{O}_k \hat{H} \rangle_{vqe}\, ,
\end{equation}
reduces to $\langle \hat{O}_k^\dag (\Delta\hat{H}) \hat{O}_k\rangle_{vqe}$ only up to zeroth order in $\beta$ through

\begin{align}
    \langle &\mathcal{L}(\hat{O}_k^\dag, \hat{H}, \hat{O}_k)\rangle_{vqe} = \langle \mathcal{L}(\hat{O}_k^\dag, \Delta \hat{H}, \hat{O}_k)\rangle_{vqe} \notag\\
    & = \langle \hat{O}_k^\dag \Delta\hat{H} \hat{O}_k \rangle_{vqe} - \beta~\text{Re}\left(\bra{v} \Delta\hat{H}\hat{O}_k^\dag \hat{O}_k \ket{\psi_{vqe}}\right) \, .
\end{align}

In deriving the third line, we have used the equality $\Delta \hat{H} \ket{\psi_{vqe}} = \beta \Delta \hat{H} \ket{u}$, which holds when subtracting the exact groundstate energy to the Hamiltonian.

\section{Measurement costs of the subspace projections}
\label[section]{Appendix_2}
A significant difference between the qEOM and qSE cost functions is that \cref{qeom_cost_function} admits an antisymmetry upon the exchange $\hat{O}_k \leftrightarrow \hat{O}_k^\dag$.
This implies that the solutions of the pseudo-eigenvalue problem appear in pairs ($\epsilon_{k0}^{qeom}, -\epsilon_{k0}^{qeom}$). Numerically, the antisymmetry translates into redundancies in the elements of the $\bm{H}_{qeom}, \bm{S}_{qeom}$ matrices compared to $\bm{H}_{qse}, \bm{S}_{qse}$ in~\cref{qse_matrix_elements}, which leads to a reduction of the required number of measurement that we summarize in \cref{table_I}. 

Being $d$ the number of expansion operators in the subspace, the number of matrix elements to compute the hermitian $\bm{H}_{qse}, \bm{S}_{qse}$ matrices in the original qSE formulation is $2 \times \frac{1}{2} \times d^2$.
Adding the de-excitation operators doubles the size of the subspace.
In the qEOM formulation, the antisymmetry reduces the number of terms to be evaluated by an additional global factor $1/2$.
However, the bigger gain comes from the rank reduction of the observables to be calculated to construct the qEOM Hamiltonian compared to the qSE one.
While qSE requires the evaluation of the two- and three-particle Reduced Density Matrices (RDMs), qEOM only requires one- and two-particle RDMs because each commutator reduces the rank by one compared to a product.

\begin{table}[!h]
    \centering
    \begin{tabular}{|c|c|c|} \hline
            Method   &         Nb observables      & Observable size   \\ \hline \hline
        qSE with TDA &     $d^2$                   & 3/2-particle RDMs \\ \hline
        qSE w/o TDA  & $(2 \times d)^2$            & 3/2-particle RDMs \\ \hline
        qEOM with TDA    & $  d^2$ & 2/1-particle RDMs \\ \hline
        qEOM w/o TDA    & $ 1/2 \times (2\times d)^2$ & 2/1-particle RDMs \\ \hline
    \end{tabular}
    \caption{Number of observables (second column) and RDM rank (third column) associated with the evaluation of the matrix elements required to build the projected $\bm{H}$, $\bm{S}$ matrices.}
    \label[table]{table_I}
\end{table}

\end{document}